\documentclass[a4paper,fleqn,twocolumn,usenatbib]{mnras}

\usepackage{mathptmx}

\usepackage[T1]{fontenc}
\usepackage{ae,aecompl}

\usepackage{graphicx}	
\usepackage{amsmath}	
\usepackage{amssymb}	
\usepackage{indentfirst}
\usepackage{verbatim}

\def\nn{\nonumber}

\def\be{\begin{equation}}
\def\ee{\end{equation}}
\def\beq{\begin{eqnarray}}
\def\eeq{\end{eqnarray}}
\def\nn{{\nonumber}}

\title[LDS with LSS induced errors]{Gravitational wave source clustering in the luminosity distance space with the presence of peculiar velocity and lensing errors}

\author[Q. Yang et al.]{
Qing Yang,$^{1}$
Bin Hu,$^{2,3}$\thanks{corresponding author: bhu@bnu.edu.cn}\\
$^{1}$College of Engineering Physics, Shenzhen Technology University, Shenzhen, 518118, China\\
$^{2}$Institute for Frontier in Astronomy and Astrophysics, Beijing Normal University, Beijing, 102206, China\\
$^{3}$Department of Astronomy, Beijing Normal University, Beijing, 100875, China\\
}

\date{Accepted XXX. Received YYY; in original form ZZZ}

\pubyear{2018}

\begin{document}
\label{firstpage}
\pagerange{\pageref{firstpage}--\pageref{lastpage}}
\maketitle

\begin{abstract}
GW number count can be used as a novel tracer of the large scale structure (LSS) in the luminosity distance space (LDS), just like galaxies in the redshift space. It is possible to obtain the $D_L-D_A$ duality relation with clustering effect. However, several LSS induced errors will contaminate the GW luminosity distance measurement, such as the peculiar velocity dispersion error of the host galaxy as well as the foreground lensing magnification. The distance uncertainties induced from these effects will degrade the GW clustering from a spectroscopic-like data down to a photometric-like data. In this paper, we investigate how these LSS induced distance errors modify our cosmological parameter precision inferred from the LDS clustering. We consider two of the next generation GW observatories, namely the Big Bang Observatory (BBO) and the Einstein Telescope (ET). We forecast the parameter estimation errors on the angular diameter distance $D_A$, luminosity distance space Hubble parameter $H_L$ and structure growth rate $f_L\sigma_8$ with a Fisher matrix method. Generally speaking, the GW source clustering data can be used for cosmological studies below $D_L<5$ Gpc, while above this scale the lensing errors will increase significantly. We find that for BBO, it is possible to constrain the cosmological parameters with a relative error of $10^{-3}$ to $10^{-2}$ below $D_L<5$ Gpc. The velocity dispersion error is dominant in the low luminosity distance range, while the lensing magnification error is the bottleneck in the large luminosity distance range. To reduce the lensing error, we assumed a $50\%$ delensing efficiency. Even with this optimal assumption, the fractional error increased to $O(1)$ at luminosity distance $D_L=25$ Gpc. The results for ET are similar as those from BBO. Due to the GW source number in ET is less than that from BBO, the corresponding results also get a bit worse.
\comment{The future gravitational wave (GW) observatories have the potential to observe millions or even more compact binary merger events during their life time. Hence, we can extend the mature techniques in the galaxy survey to the GW analysis. As one of the robust cosmology probes, the number count clustering can tell us fruitful information about the cosmic density field and velocity field. GW number count can be used as a novel tracer of the large scale structure (LSS) in the luminosity distance space (LDS), just like galaxies in the redshift space. Thanks to the clustering effect, it is possible to obtain the $D_L-D_A$ duality relation. However, several LSS induced errors will contaminate the GW luminosity distance measurement, such as the peculiar velocity dispersion error of the host galaxy as well as the foreground lensing magnification. The distance uncertainties induced from these effects will degrade the GW clustering from a spectroscopic-like data down to a photometric-like data. In this paper, we investigate how these LSS induced distance errors modify our cosmological parameter precision inferred from the LDS clustering. We consider two of the next generation GW observatories, namely the Big Bang Observatory (BBO) and the Einstein Telescope (ET). We forecast the parameter estimation errors on the angular diameter distance $D_A$, luminosity distance space Hubble parameter $H_L$ and structure growth rate $f_L\sigma_8$ with a Fisher matrix method. Generally speaking, the GW source clustering data can be used for cosmological studies below $D_L<5$ Gpc, while above this scale the lensing errors will increase significantly. We find that  for BBO, it is possible to constrain the cosmological parameters with a relative error of $10^{-3}$ to $10^{-2}$ below $D_L<5$ Gpc. The velocity dispersion error is dominant in the low luminosity distance range, while the lensing magnification error is the bottleneck in the large luminosity distance range. To reduce the lensing error, we assumed a $50\%$ delensing efficiency. Even with this optimal assumption, the fractional error increased to $O(1)$ at luminosity distance $D_L=25$ Gpc. The results for ET are similar as those from BBO. Due to the GW source number in ET is less than that from BBO, the corresponding results also get a bit worse. }
\end{abstract}




\section{Introduction}
\label{sec:Intro}
The era of gravitational-wave (GW) astronomy has begun with the detection of the first GW signal GW150914 from the merger of binary black holes (BBHs) by the Advanced LIGO interferometers \cite{LIGOScientific:2016aoc}. More than ninety GW events \cite{LIGOScientific:2021djp} has been discovered in the following years after the first detection, including one with follow-up electromagnetic signals GW170817 \cite{LIGOScientific:2017vwq}. It is expected that with this new observational window, we can explore the nature of gravity, and also extract useful astronomy and cosmological informations.

To learn about the universe and its cosmic expansion history we need to measure both distances and redshifts. One shortcome of GW is that it does not provide direct information about redshift. If EM counterpart can be identified, redshift of the source can be directly extracted. However, the probability for finding EM counterparts to the stellar mass BBH mergers is almostly theoretically prohibited. Various methods has been developed to circumvent this need for EM follow-ups. For example, utilizing the anisotropies of compact binaries originated from the large-scale structure \cite{Namikawa:2015prh}; the cross-correlation of gravitational wave standard sirens and galaxies \cite{Oguri:2016dgk}; \cite{Nair:2018ign}; \cite{Mukherjee:2018ebj}; \cite{Zhang:2018ekk}, the statistical information of galaxy redshifts \cite{Chen:2017rfc,Fishbach:2018gjp,Gray:2019ksv,Wang:2020dkc,Zhu:2021aat}, the mass distribution function of binary black hole sources \cite{Farr:2019twy}; \cite{You:2020wju}, and other methods \cite{Seto:2001qf}; \cite{Messenger:2011gi}; \cite{DelPozzo:2015bna}, {\it etc.}

During recent years, several updated GW detectors have been proposed around the world. The future spaceborne GW observatories and the third-generation ground-based detectors will achieve unprecedented sensitivity in a broad frequency range. The Einstein Telescope (ET) is expected to have a sensitivity 10 times better than the current second-generation instruments, and may have a detection rate of $10^5$ events per year  \cite{Punturo:2010zz}; \cite{TP-toolbox-web}. The spaceborne GW observatory, such as the Laser Interferometer Space Antenna (LISA) \cite{Audley:2017drz}, Taiji \cite{Hu:2017mde} and TianQin \cite{TianQin:2015yph}, will open up a $10^{-4}\sim 1$ Hz window to the gravitational universe. The Big Bang Observer (BBO), which is a proposed successor to LISA, mainly focuses on the observation of gravitational waves from physical precesses shortly after the Big Bang. It will be able to detect almost {\it all} the GW signals from compact binaries in the universe, and may have a detection rate of $10^{7}$ events per year \cite{Cutler:2005qq}; \cite{Cutler:2009qv}.

With more and more GW events to be detected by the future GW observatories, the large scale structure (LSS) can be traced in the luminosity-distance space (LDS), just as what has already been done in the redshift surveys of galaxies \cite{Zhang:2018nea}. Hence, one way to solve the issues associated with dark sirens is to construct a 3D map of GW sources in LDS. Via the anisotropic clustering measurement, one can obtain the luminosity distance ($D_L$) as well as angular diameter distance ($D_A$) in different redshift bins. 
By utilizing these $D_L-D_A$ relation, one can extract the LSS formation and evolution law in LDS. It plays essentially the same role as the standard redshift-distance duality. 
Several works has been done in this direction: \cite{Zhang:2018nea,Libanore:2020fim,Palmese:2020kxn,Zhang:2021tdt,Libanore:2021jqv,Namikawa:2020twf,Yu:2020agu}.

In this paper, we are going to constrain the cosmological expansion history as well as LSS growth rate in LDS, {\it ie.} the angular diameter distance $D_A$, luminosity distance space Hubble parameter $H_L$ and the growth rate $f_L\sigma_8$. 
We adopt a Fisher matrix analysis to give the measurement errors for each of these parameters. As for the experimental configurations, we explore the capability of ET and BBO.  
In particular, we investigate some major systematics to the luminosity distance measurement, namely the weak lensing magnification along the line of sight \cite{PhysRevD.81.124046}, as well as the peculiar velocities in the low redshift regime \cite{Gordon:2007zw}. 
We consider both of these effects and see how they will downgrade the cosmological parameter constraining ability of the GW LDS measurements.

The rest of the paper is organized as follows. In section \ref{sec:Methods}, we will first briefly overview the physical picture of luminosity distance space and the anisotropy power spectrum in LDS. And then, we will introduce some details of our analysis, including the source of luminosity distance measurement errors, the event rate, and the tomographic redshift distribution of the GW sources. 
In Section \ref{sec:res}, we present our forecasted result by considering BBO and ET. Section \ref{sec:con} is devoted to summaries and discussions.

\section{Methods}
\label{sec:Methods}

GW signals do not directly provide the redshift information. However, since the amplitude of the GW signal is inverse proportional to the luminosity distance to the source, and masses and orbital inclination can be extracted from frequency evolution and relativistic amplitude, the GW signals can provide a direct and absolute measurement of the luminosity distance. 
A map of large scale structure in luminosity distance space could be constructed from GW signals when the number density of the GW sources reached some threshold. In this section we will first briefly introduce the physical picture of statistical anisotropic power spectrum in the luminosity distance space, which was firstly proposed by \cite{Zhang:2018nea}. 

Due to the presence of peculiar velocity, it will contaminate the background recession velocity and hence bias the luminosity distance.   
Up to the first order in the peculiar velocity, the observed luminosity distance reads
\beq
D_L^{\rm obs}\approx D_L(1+2\vec{v}\cdot \hat{n})\;,
\eeq
where $D_L$ is the luminosity distance in the unperturbed background, $\vec{v}$ is the peculiar velocity of the source, and $\hat{n}$ is the line of sight unit vector. 
This effect distorts the matter distribution pattern in LDS w.r.t. the one assuming the unperturbed background. 
It resembles the redshift space distortion (RSD) in the galaxy spectrographic survey. 
The resulted power spectrum reads 
\beq\label{PSinLDS}
P^{\rm LDS}(k_{\perp}, k_{||})=P_m(k)\bigg{(}1+\frac{f_L}{b_g}\mu^2\bigg{)}^2F(k_{||})\;,
\eeq
where $k_{\perp}(k_{||})$ represents the wave vector perpendicular (parallel) to the line of sight, $\mu$ is the cosine of the angle between the line of sight direction and the peculiar velocity of the source with  $\mu\equiv k_{||}/k$ and $k\equiv \sqrt{k^2_{\perp}+k^2_{||}}$, $b_g$ is the density bias of GW host galaxies, and $F(k_{||})$ describe the finger of god (FoG) effect. Eq.~\eqref{PSinLDS} has similar structure as the formula in RSD, except for the pre-factor in the second term where
\beq
f_L=\Bigg{(}\frac{2D_L/(1+z)}{d(D_L)/dz}\Bigg{)}\times f\;,
\eeq
and $f$ is the standard dimensionless growth rate defined as
\beq
f=\frac{a}{D_1}\frac{dD_1}{da}=\frac{1}{aHD_1}\frac{dD_1}{d\eta}\;,
\eeq
where $a$ is the scale factor, $\eta$ is the conformal time, and $D_1$ is the linear growth factor.
$f_L$ is zero at $z=0$, and increases monotonically with redshift. As a result, the peculiar velocity induced distortion effect in LDS is negligible at small redshift, but become larger than that in redshift space at around $z=1.7$.

In this paragraph, we will introduce the details of the methods which we used to constrain the cosmological parameters.
We employ a Fisher matrix analysis for 3 parameters, namely $D_A$, $H_L$, and $f_L\sigma_{8}$. 
We adopt the same method as well as the same parameters as those in  \cite{Zhang:2018nea}. 
However, the difference lies in we considered the luminosity distance errors induced by the lensing magnification and peculiar velocity.  
We would like to have a direct comparisions with those without considering these unavoidable uncertainties. 
Under the Gaussian statistical assumptions, the Fisher matrix is given by 
\beq\label{Fisher}
F_{\alpha\beta}=\sum_{\textit{\textbf{k}}} \frac{\partial P^{\rm LDS}(k,\mu, z)}{\partial \lambda_{\alpha}}\frac{\partial P^{\rm LDS}(k,\mu, z)}{\partial \lambda_{\alpha}}\frac{1}{P^{\rm LDS}(k,\mu.z)^2}V_{\rm eff}(k)\;.
\eeq
Here the sum is over different $\textit{\textbf{k}}$ bins, $P^{\rm LDS}$ is the statistically anisotropic power spectrum given in Eq.~\eqref{PSinLDS}, and $\lambda_\alpha$ stands for the cosmological parameters. For the matter power spectrum in Eq.~\eqref{PSinLDS}, we simply use the empirical formula in \cite{White:1996pz}, {\it ie.}
\beq
P_m(k)=\frac{2\pi^2k}{H_0^4}\delta^2_H(k)T^2(k)\;,
\eeq
where $\delta_H$ is the density fluctuation and $T(k)$ is the transfer function.
$V_{\rm eff}$ in Eq.~\eqref{Fisher} is the effective survey volume for a certain $\textit{\textbf{k}}$ shell, which is given by \cite{Feldman:1993ky,Hamilton:1997kv,Tegmark:1997rp}
\beq
V_{\rm eff}^{(i)}(k)=\int\bigg{(}\frac{\bar{n}_{i}(z)P(k,z)}{W_{||}^{-2}(k)W_{\perp}^{-2}(k)+\bar{n}_{i}(z)P(k,z)}\bigg{)}^2d^3\textit{\textbf{r}}\;,
\label{eq:Veff}
\eeq
where $W_{||}$ and $W_{\perp}$ are window functions in the parallel and perpendicular directions. $\bar{n}_{i}(z)$ is the source number density in the $i$-th ``photo-$z$'' bin, which we will define later. 
Since we are interested at large scale, we can take $W_{\perp}=1$ to a good approximation. However, since we are investigating the distortion effect induced by the peculiar velocity field (a sub-leading order effect), we have to keep the same order term in the window function along the parallel direction. Hence, we have $W_{||}=\exp(-k^2\chi^2\sigma^2_{\log D}/2)$, among which $\chi$ is the conformal distance, and $\sigma_{\log D}$ is the distance measurement error in the logarithmic scale. 

Besides of the experimental error of the luminosity distance which has already been considered in \cite{Zhang:2018nea}, in this paper we also include some of the unavoidable distance uncertainties which are generated during the propagation along the line of sight. Among these, two of the most significant effects are the weak lensing magnification in the high redshift and the peculiar velocity in the low redshift. We adopt the following fitting formula \cite{PhysRevD.81.124046,Kocsis_2006,PhysRevLett.99.081301} for the distance errors due to lensing magnification $\sigma^{\rm lens}_{D_L}$ and that due to velocity dispersion $\sigma^{\rm pv}_{D_L}$
\beq
\label{eq:lens1}
\sigma_{D_L}^{\rm lens}(z)=D_L(z)\times C_l \bigg{(}\frac{1-(1+z)^{-\beta_l}}{\beta_l}\bigg{)}^{\alpha_l}\;,\\
\label{eq:pv1}
\sigma_{D_L}^{\rm pv}(z)=D_L(z)\times \bigg{(}1-\frac{c(1+z)^2}{H(z)D_L(z)}\bigg{)}\frac{\sqrt{\langle v^2\rangle}}{c}\;,
\eeq
where $C_l=0.066, \beta_l=0.25, \alpha_l=1.8$ and $\sqrt{\langle v^2\rangle}=500~km/s$, which is the peculiar velocity root mean square of the host galaxy with respect to the Hubble flow. For the intrinsic measurement uncertainties, we consider four cases with $\sigma_{\log D}^{\rm GW}=0.001, 0.005, 0.01, 0.02$, which may well cover the typical $D_L$ uncertainty in the whole redshift range. Hence, the total uncertainty is given by
\beq
\sigma_{D_L}^{\rm tot}=\sqrt{(\sigma^{\rm GW}_{D_L})^2+(\epsilon\sigma^{\rm lens}_{D_L})^2+(\sigma^{\rm pv}_{D_L})^2}\;.
\eeq
The coefficient $\epsilon$ in front of $\sigma^{\rm lens}_{D_L}$ denotes for the delensing efficiency. 
This is due to the fact that lensing is dominant error on the luminosity distance estimation in the high redshift, in order to utilise the GW data for precision cosmology purpose, we have to do the delensing operation to the original GW data. However, given the experience from CMB \cite{SimonsObservatory:2018koc}, we know that it is technically very challenging. 
For GW delensing, we can use the synthetic observation of wide shallower weak lensing shear measurement ({\it eg.} 20 galaxy/arcmin$^2$) and small deeper weak lensing flexion measurement ({\it eg.} 500 galaxy/arcmin$^2$) \cite{Shapiro:2009sr,Hilbert:2010am}. In this paper, we adopt an optimal choice of delensing efficiency, namely a $50\%$ delensing, which is the similar level as the stage-IV CMB experimental capability \cite{SimonsObservatory:2018koc}.   

Next, let us introduce the true average number density of GW source $\bar n_{\rm GW}$. It is different from $\bar{n}_{i}(z)$ in Eq. (\ref{eq:Veff}). The latter is the observed averaged number density in the $i$-th ``photo-$z$'' bin. Here, we borrow the phrase ``photo-$z$'' from the galaxy survey to describe the luminosity distance measurement uncertainty in GW data. The ``observed'' photo-$z$ can be computed from the observed luminosity distance by assuming some fiducial distance-redshift duality.
The true density and the ``photo-$z$'' density are related via
\beq
\bar{n}_{i}(z)=\int_{z^i_{p, {\rm low}}}^{z^i_{p,{\rm up}}}~\bar n_{\rm GW}(z)p(z_p|z)~dz_p\;,
\eeq
where $z_p$ is the measured ``photo-$z$'', $p(z_p|z)dz_p$ describes the true $z$ distribution in a certain ``photo-$z$'' interval.  
We consider both BH-BH and NS-NS mergers that may be detected in a ten year period by ET and BBO.   
For BBO, $\bar n_{\rm GW}$ is dominated by NS-NS events, we adopt a model in \cite{Cutler:2009qv} to describe the redshift evolution the NS-NS merger rate
\beq
\bar n_{\rm GW}(z)=n_0\times r(z)\;,
\eeq
where $r(z)$ is the time evolution factor which is given by the following linear fit formula
\beq
 \begin{split}
 r(z)= \left \{
 \begin{array}{ll}
 1+2z\;,                    & z\le 1\\
\frac{3}{4}(5-z)\;,     & 1\le z\le 5\\
 0\;,                                & z\ge 5
 \end{array}
 \right.
 \end{split}
 \eeq
The local merger rate $r_0$ for BBO is assumed to be $1.54\times 10^{-6}~\rm{Mpc^{-3}yr^{-1}}$.
For ET, we adopt an interpolation formula given in \cite{Libanore:2020fim}
\beq
\frac{d^2N_m}{dzd\Omega}=2\bigg{[}A\exp \bigg{(}-\frac{(z-\bar{z})^2}{2\sigma^2}\bigg{)}\bigg{]}
\bigg{[}\frac{1}{2}\bigg{(}1+{\rm erf}(\frac{\alpha(z-\bar{z})}{\sigma^2\sqrt{2}})\bigg{)}\bigg{]}\;,
\eeq
where $A=10^{3.22}, \bar{z}=0.37, \sigma^2=1.42, \alpha=5.48$ for BH-BH merger events, and $A=10^{3.07}, \bar{z}=0.19, \sigma^2=0.15, \alpha=0.8$ for NS-NS merger events.

We adopt the tomographic method for measuring the LDS power spectrum. We divide the survey depth range, $D_L\in(0,30)$Gpc, into three bins. As for comparison, \cite{Zhang:2018nea} divide the same luminosity range into six bins. The differences are due to in our case we take into account the lensing magnification and peculiar velocity uncertainties in the luminosity distance estimation. From Eq. (\ref{eq:lens1}) and (\ref{eq:pv1}), one can see that the corresponding errors range from a few percent to twenty percent, especially for the lensing magnification in the high redshifts. These unavoidable errors degrade the $D_L$ measurement from the spectroscopic-like survey to the photometric-like survey. In details, for BBO we choose the redshift bins as $z\in(0-1.54)$, $(1.54-2.40)$, $(2.40-3.70)$; for ET, we have $z\in(0-0.71)$, $(0.71-1.36)$, $(1.36-6)$. Each bin contains comparable amounts of GW events. 

\begin{figure}
\centering
\includegraphics[width=0.4\textwidth]{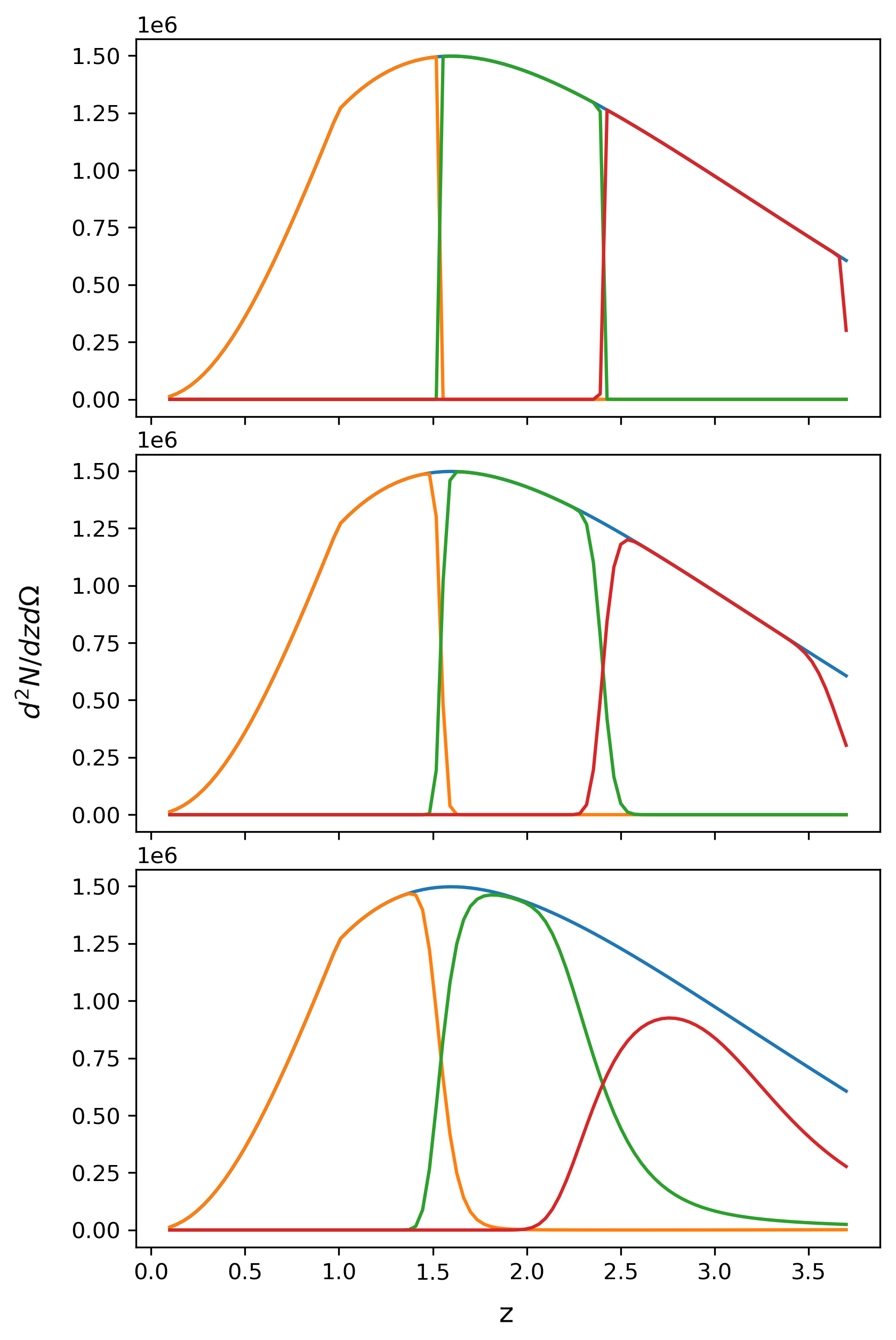}
\caption{\small The true redshift distributions in different tomographic bins with BBO experimental configuration. Top panel: $D_L$ uncertainty from GW intrinsic measurement error only; Middle panel: $D_L$ uncertainty from GW intrinsic measurement error plus peculiar velocity error; Bottom panel: $D_L$ uncertainty from GW intrinsic measurement error plus lensing magnification error. All three panels assume the GW intrinsic measurement error level, $\sigma_{\log D_L}^{\rm GW}=0.001.$}\label{BBObin}
\end{figure}

\begin{figure}
\centering
\includegraphics[width=0.4\textwidth]{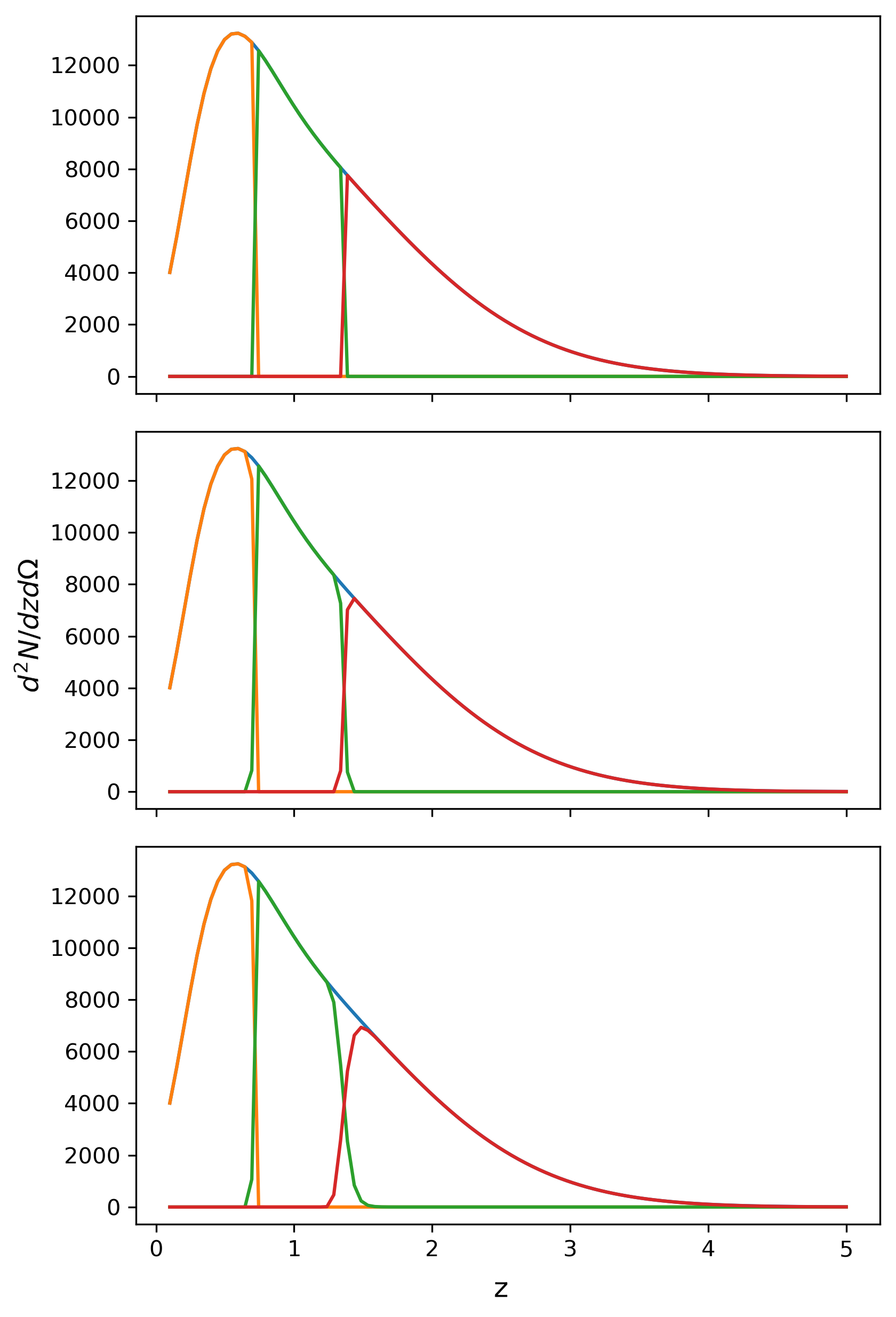}
\caption{\small The true redshift distributions in different tomographic bins with ET experimental configuration. Top panel: $D_L$ uncertainty from GW intrinsic measurement error only; Middle panel: $D_L$ uncertainty from GW intrinsic measurement error plus peculiar velocity error; Bottom panel: $D_L$ uncertainty from GW intrinsic measurement error plus lensing magnification error. All three panels assume the GW intrinsic measurement error level, $\sigma_{\log D_L}^{\rm GW}=0.001.$}\label{ETbin}
\end{figure}

Hence, one can adopt the conventional procedure in the photometric galaxy survey in our GW analysis.  
We take a simple assumption that the deduced redshift has a Gaussian scatter given the true redshift, just like the photometric redshift distributions given the true redshift in photometric galaxy surveys \cite{Ma:2005rc}; \cite{Yao:2017dnt}
\beq
p(z'|z)=\frac{1}{\sqrt{2\pi\sigma_z(1+z)}}\exp \bigg{[}-\frac{(z-z'-\Delta^i_z)^2}{2(\sigma_z(1+z))^2}\bigg{]}\;,
\eeq
where $z$ is the true redshift, $z'$ is the photo-$z$ redshift, $\sigma_z$ is the redshift uncertainty induced from the luminosity distance measurement error. In Fig.~\ref{BBObin} (for BBO) and Fig.~\ref{ETbin} (for ET), we show the true $z$ distributions for the three tomographic bins with the different luminosity distance uncertainties, and thus different photo-$z$ scatter $\sigma_z$.

One can see that, if we only consider the intrinsic measurement error as shown in the top panel of Fig.~\ref{BBObin} and Fig.~\ref{ETbin}, there are negligible overlap among different redshift bins. This is similar to the case of spectroscopic redshift in galaxy survey. Once we include the statistical errors from the peculiar velocity (Middle panel) and lensing magnification (Bottom panel), the true redshifts are scattered back and forth in and among the bins. The true redshift distributions of different photo-$z$ bins overlap significantly. 
We will see how this will affect the cosmological parameter uncertainties is the next section.

\section{Results}
\label{sec:res}

In this section we present our Fisher matrix results for the measurement errors on $D_A$, $H_L$ and $f_L\sigma_8$ . We assume a ten year observation time for both BBO and ET. We show the results with different types of luminosity distance errors.

\begin{figure*}
\centering
\begin{minipage}{0.32\textwidth}
\includegraphics[width=1\textwidth]{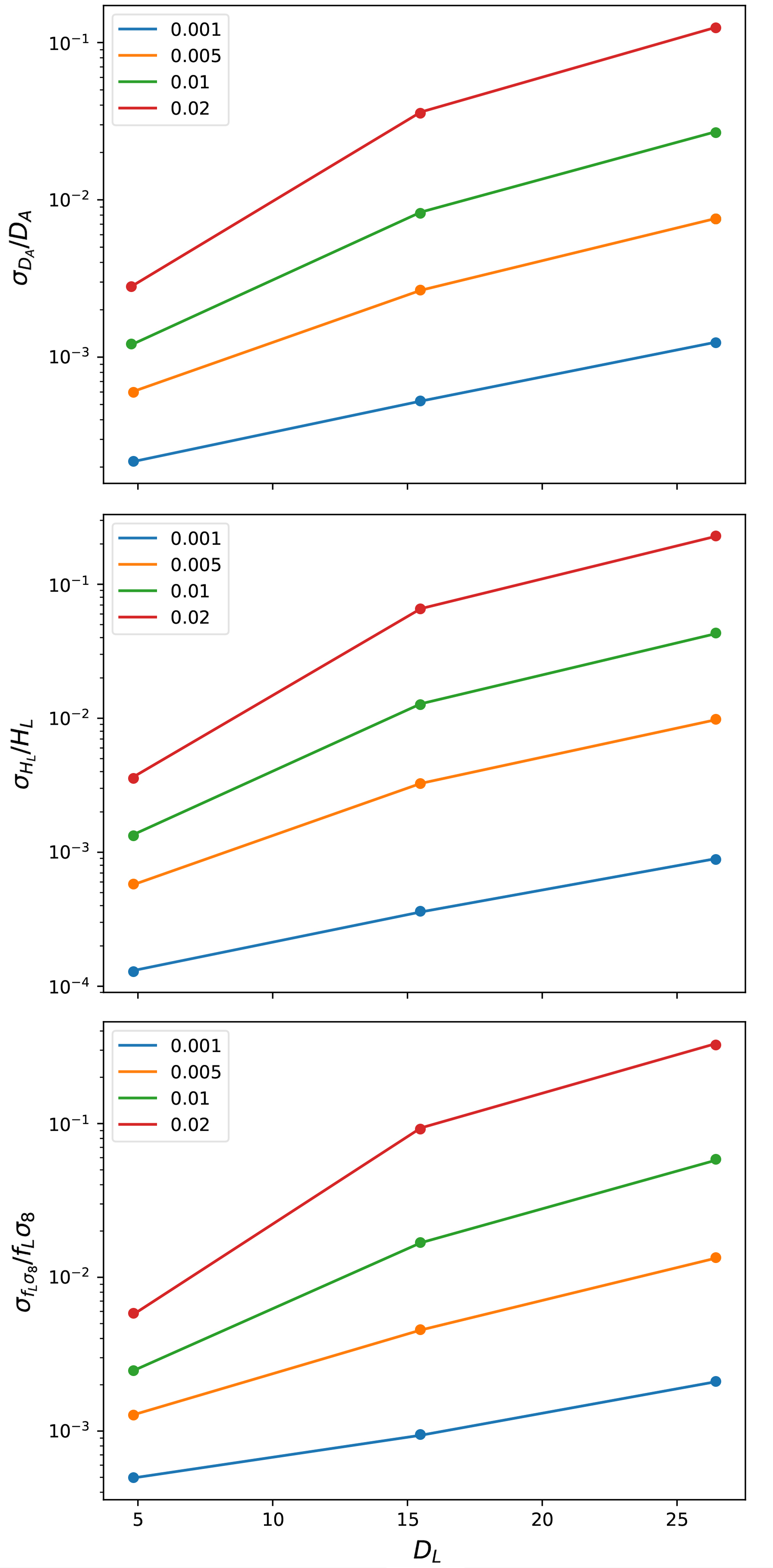}
\end{minipage}
\begin{minipage}{0.32\textwidth}
\includegraphics[width=1\textwidth]{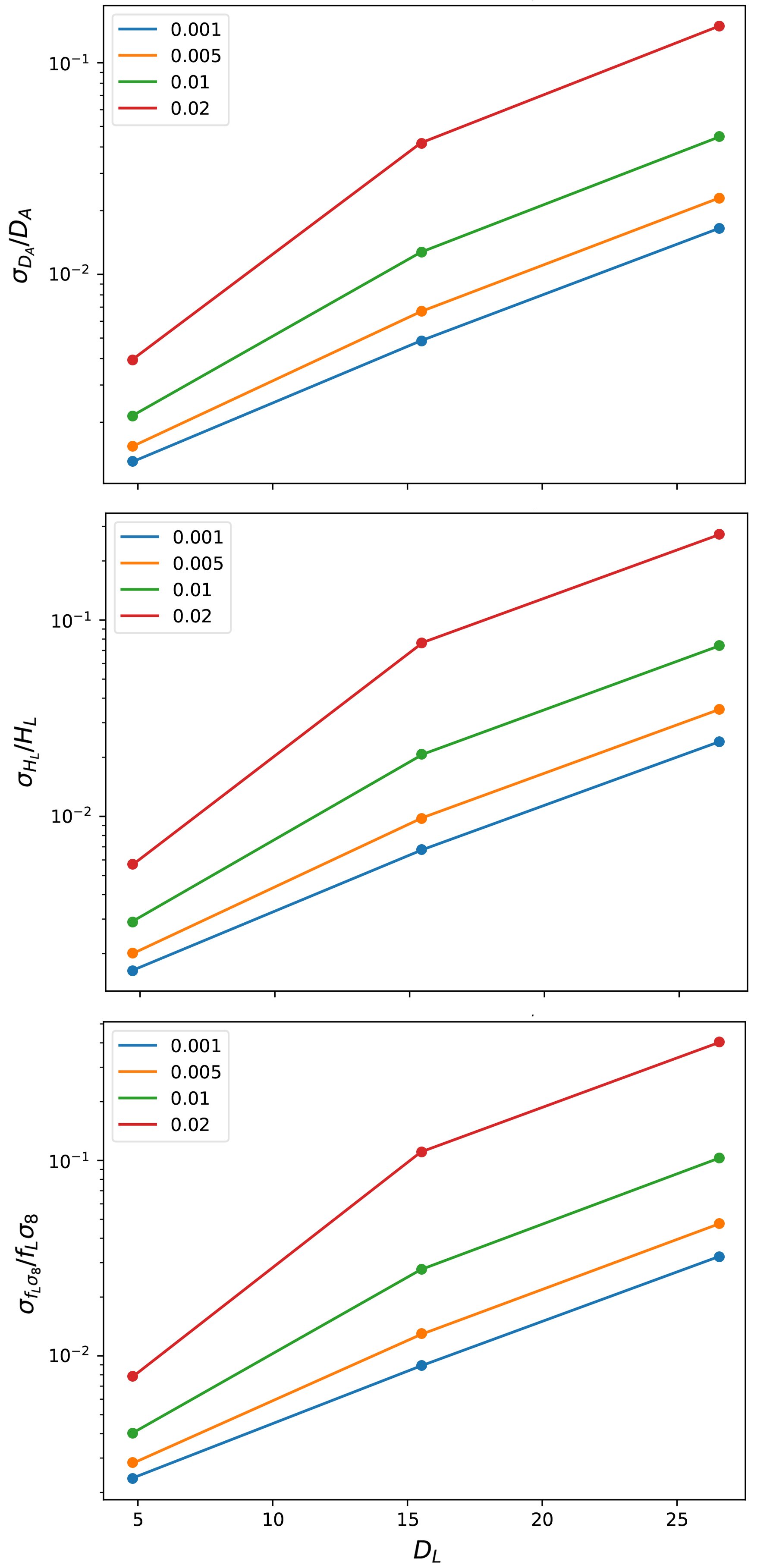}
\end{minipage}
\begin{minipage}{0.32\textwidth}
\includegraphics[width=1\textwidth]{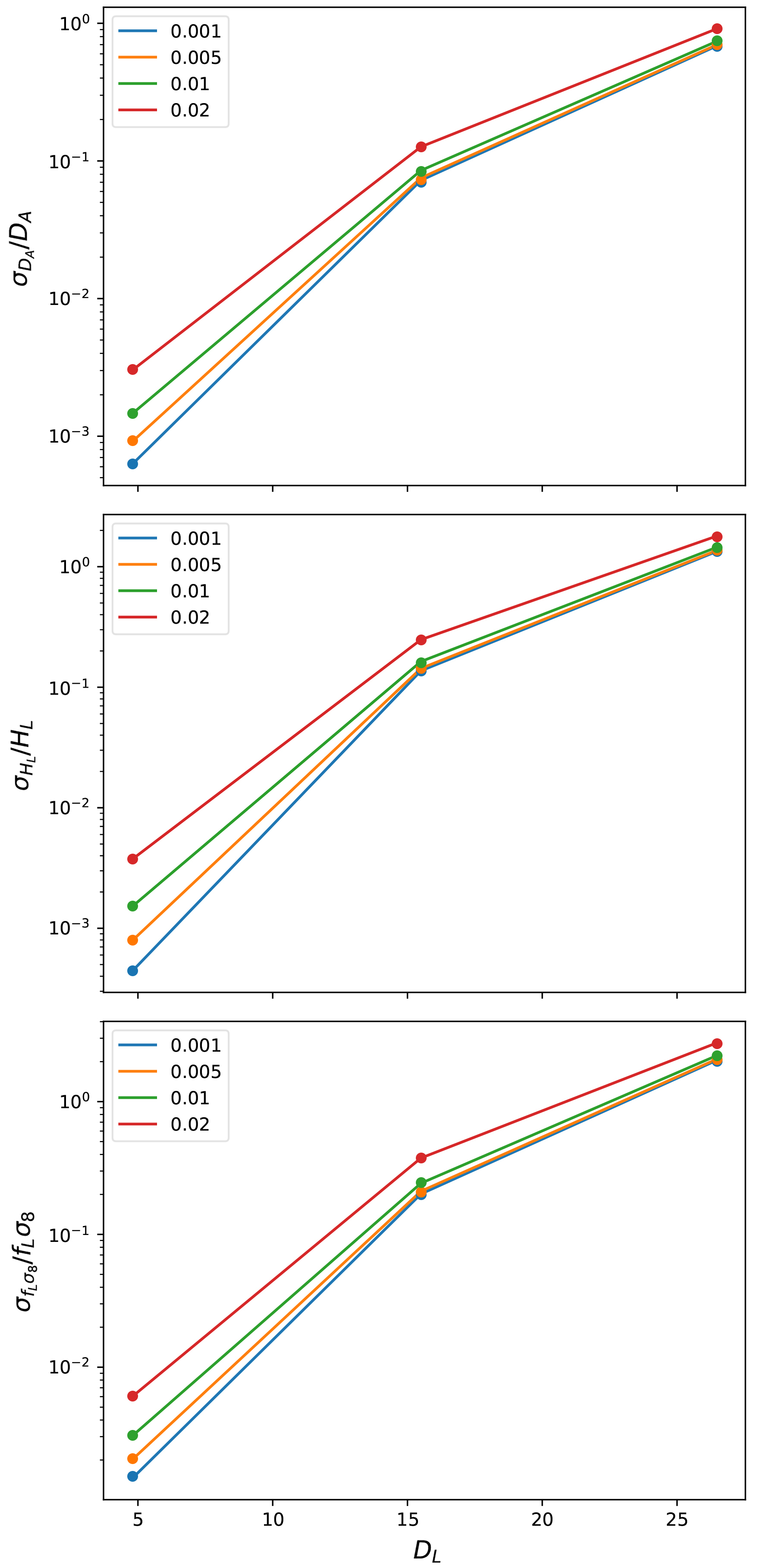}
\end{minipage}
\caption{\small The forecasted errors on three cosmological parameters from three tomographic bins, assuming a ten-year observation period for BBO. The red, green, yellow and blue curves denote for different luminosity distance measurement errors, namely $\sigma^{\rm GW}_{\log D_L}=0.001, 0.005, 0.01$ and $0.02$. Left panel: $D_L$ measurement uncertainty is from the intrinsic GW observations error only; Middle panel: $D_L$ measurement uncertainty is from the intrinsic GW observations and peculiar velocity dispersion errors; Right panel: $D_L$ measurement uncertainty is from the intrinsic GW observations and gravitational lensing errors.}\label{BBOresult}
\end{figure*}

\begin{figure*}
\centering
\begin{minipage}{0.32\textwidth}
\includegraphics[width=1\textwidth]{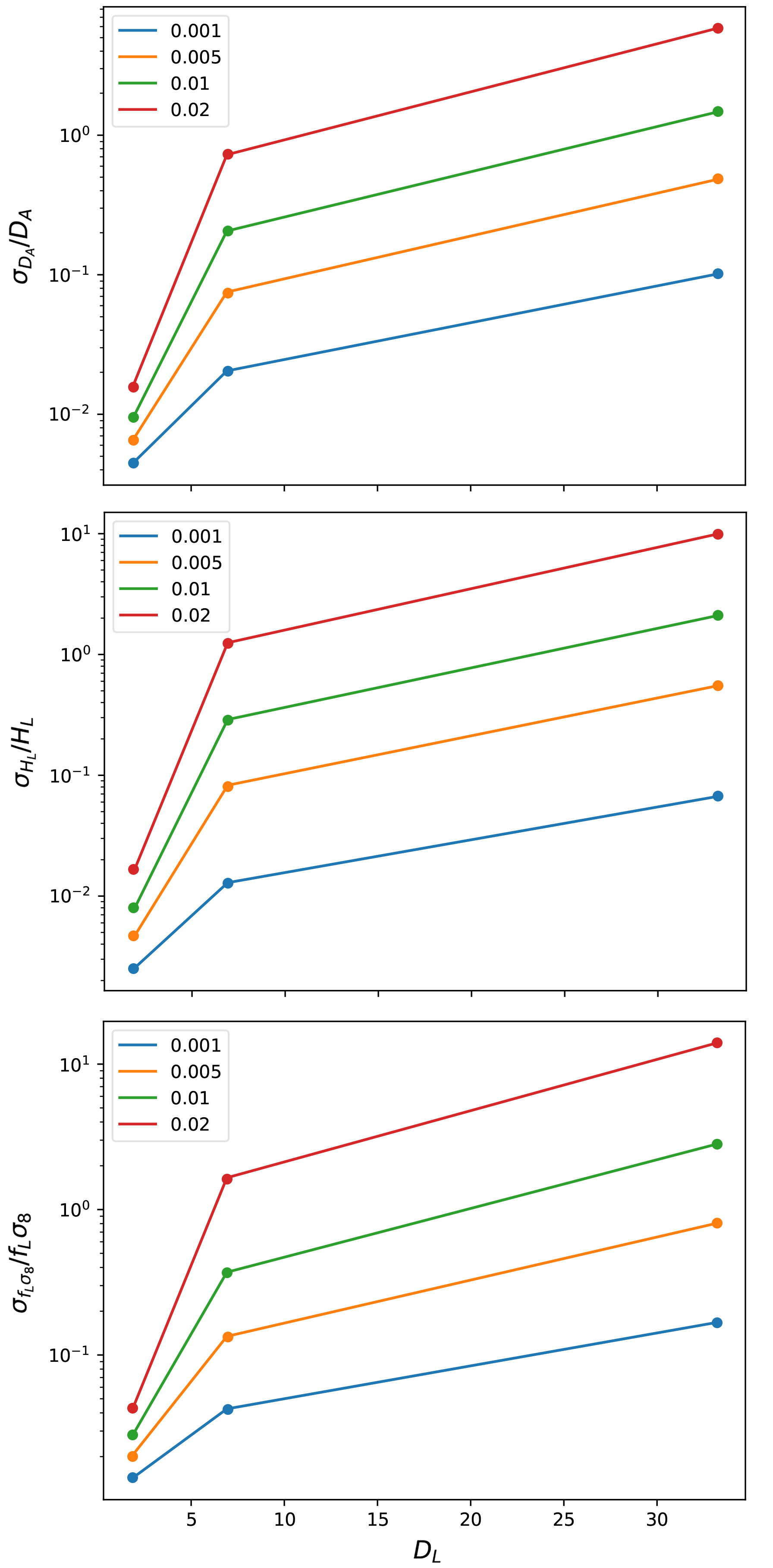}
\end{minipage}
\begin{minipage}{0.32\textwidth}
\includegraphics[width=1\textwidth]{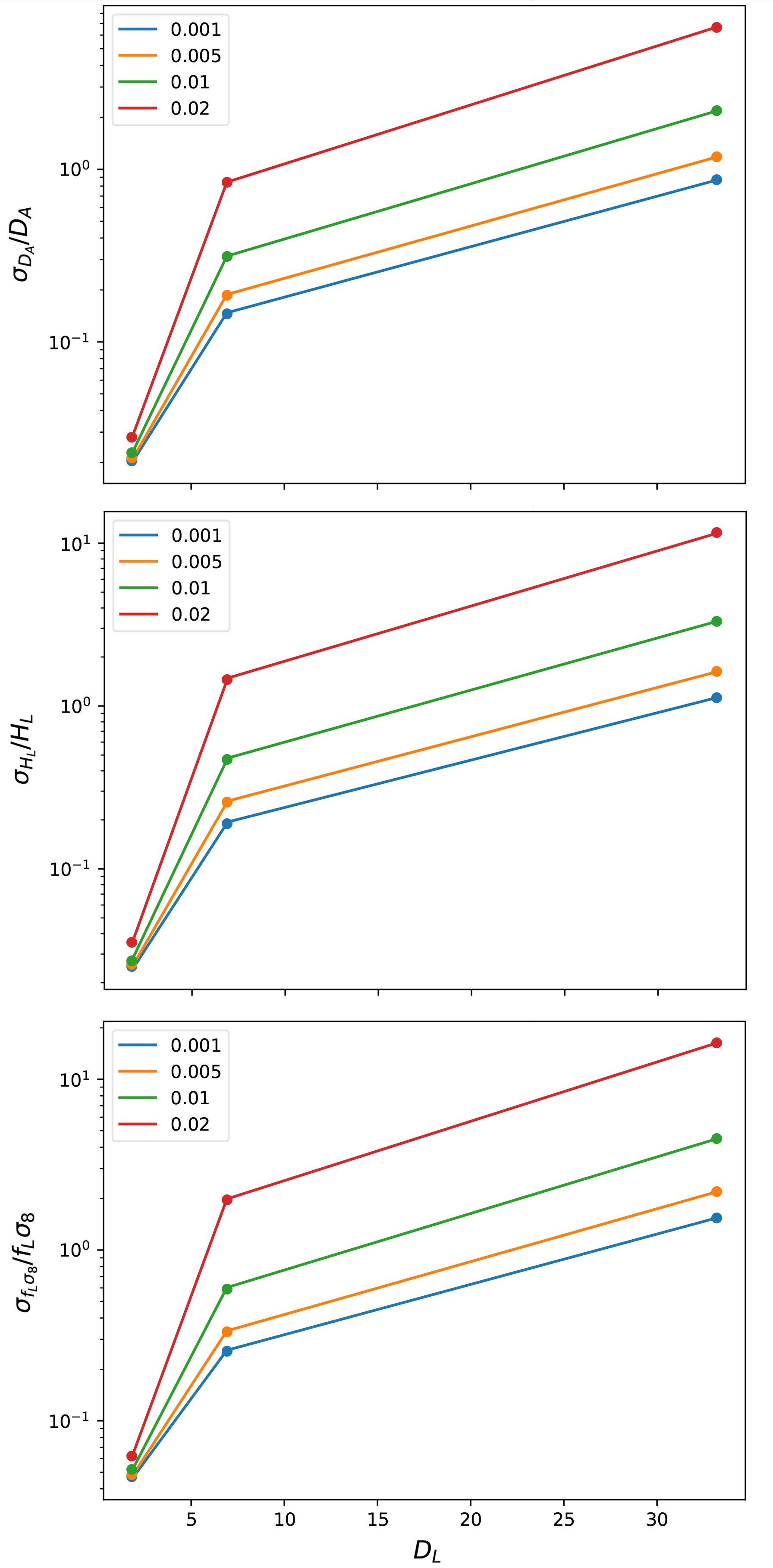}
\end{minipage}
\begin{minipage}{0.32\textwidth}
\includegraphics[width=1\textwidth]{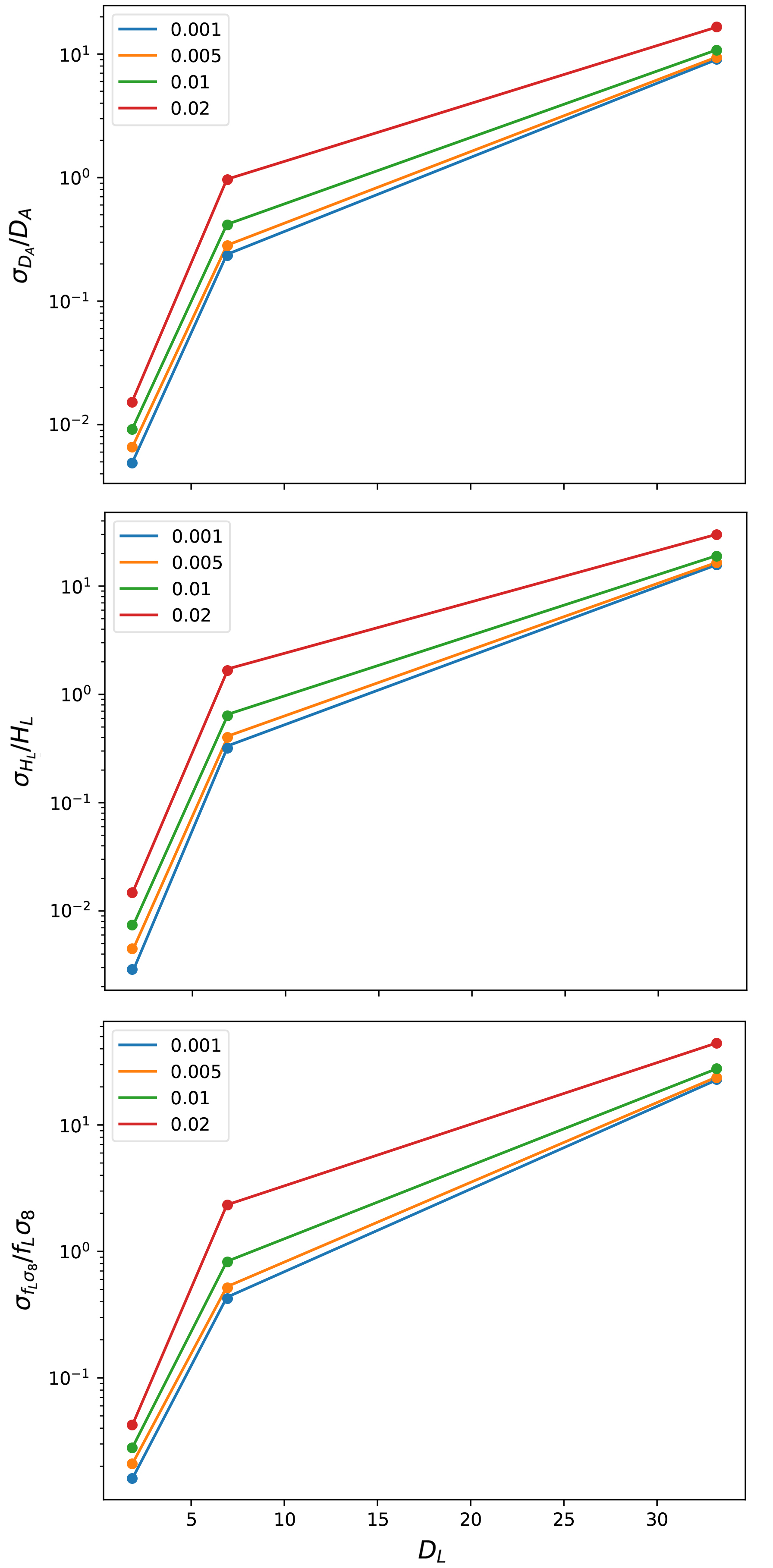}
\end{minipage}
\caption{\small The forecasted errors on three cosmological parameters from three tomographic bins, assuming a ten-year observation period for ET. The red, green, yellow and blue curves denote for different luminosity distance measurement errors, namely $\sigma^{\rm GW}_{\log D_L}=0.001, 0.005, 0.01$ and $0.02$. Left panel: $D_L$ measurement uncertainty is from the intrinsic GW observations error only; Middle panel: $D_L$ measurement uncertainty is from the intrinsic GW observations and peculiar velocity dispersion errors; Right panel: $D_L$ measurement uncertainty is from the intrinsic GW observations and gravitational lensing errors.}\label{ETresult}
\end{figure*}

In Fig.~\ref{BBOresult}, we show our result for BBO. The three panels represents for the cases with different luminosity errors, {\it ie.}, only the intrinsic measurement error; intrinsic error plus the peculiar velocity error; intrinsic error plus lensing magnification error. 
It is clearly seen from the figure that, the uncertainty for the cosmological parameters increases with increasing redshift. 
This is due to the decrement of GW event detection rate at high redshifts. 
In the case with only $D_L$ intrinsic measurement uncertainty, {\it ie.} the left panel of Fig.~\ref{BBOresult}, the cosmological parameters can be detected with high precision. 
The fractional errors for all the three parameters are below $10^{-3}$ for the case with $\sigma^{\rm GW}_{\log D_L}=0.001$. For the worst case, namely $\sigma^{\rm GW}_{\log D_L}=0.02$, the estimated fractional errors can be degraded into $10^{-1}$. 
In the middle and right panel of Fig.~\ref{BBOresult}, we see that once the LSS induced luminosity distance measurement uncertainties are included, the errors on the cosmological parameters increase significantly. 
In the case with velocity dispersion errors, the fractional errors for cosmological parameters stay in the range of a few $10^{-3}$ to $10^{-1}$. 
This is still an acceptable results. 
However, once the lensing uncertainty is included, the fractional errors increase dramatically in the high redshift. It reaches about order $O(1)$ in the third luminosity bin which is centered at $D_L=25$Gpc. In the first and second luminosity bins, the results with lensing errors are similar as those with velocity dispersion errors. 
In Fig.~\ref{ETresult}, we show our result for ET.
The results have similar tendency as BBO. 
However, since the detection rate for ET is much smaller than that for BBO, the fractional error is about 1 to 2 orders of magnitudes worse than that of BBO. 
For the case with only the intrinsic $D_L$ measurement errors, {\it ie.} the left panel of Fig.~\ref{ETresult}, the fractional errors in the first luminosity bin ranges from 
a few $10^{-3}$ to $10^{-2}$ for different intrinsic luminosity errors. The corresponding errors in the second luminosity bin reach the level of $10^{-2}$ to 1. The former is for  $\sigma^{\rm GW}_{\log D_L}=0.001$ and the latter is for $\sigma^{\rm GW}_{\log D_L}=0.02$. 
For the case after adding the velocity dispersion to the luminosity distance uncertainty, the results in the second and third luminosity bins (middle panel of Fig.~\ref{ETresult}) are almost unchanged compared with the one with only the intrinsic measurement error. The results in the first bin are about a factor of few worse than those shown in the left panels. This is different from the BBO result. For BBO, the intrinsic luminosity distance measurement error is small, hence the results are sensitive to the addition of velocity dispersion. However, for ET, the addition of velocity dispersion error does not change the results significantly.
For the case with the inclusion of lensing magnification errors, as shown in the right panel of Fig.~\ref{ETresult}, the data in the third luminosity bin is completely useless for the cosmological purpose. For instance, the fractional error of $f_L\sigma_8$ has increased to $5\times 10^1$. 
On the other side, since the lensing effect is still sub-dominant in the low redshift, the cosmological results from the first luminosity bin is still comparable w.r.t. the galaxy survey method.

\section{Conclusions}
\label{sec:con}

Unlike galaxy surveys, GW events are observed in the luminosity distance space (LDS) rather than the redshift space. With the accumulation of the GW events, one can use this data as a tracer of the large scale structure and hence study its cosmology implication.  
In this work, we investigate the possibility of using GW source clustering data in the luminosity distance space to constrain the cosmological parameters. In particular, we study the contamination effects from weak lensing magnification and peculiar velocity dispersion uncertainties. In order to qualify the cosmology study requirement of using number counts, the GW source events need to be accumulated to or above the level of 1 million. To satisfy this condition, we consider the future spaceborne GW observatories, such as Big Bang Observatory (BBO), and the third-generation ground-based observatory, such as Einstein Telescope (ET). For each of these two experiments, we assumed a ten-year observation period. For BBO, one can accumulate $10^8$ BH-BH and NS-NS merger events within 10 years; for ET, this number is about  $10^6$. 

We adopted a Fisher matrix analysis to constrain several cosmological parameters, {\it ie.} the angular diameter distance $D_A$, luminosity distance space Hubble parameter $H_L$, and the linear growth rate $f_L\sigma_8$. 
We considered several different sources of luminosity distance measurement uncertainty, {\it ie.} velocity dispersion, gravitational lensing as well as the intrinsic uncertainty from GW detection. We forecasted the fractional errors of the cosmological parameters with different $D_L$ error budget. We showed how they affect the cosmological parameter constraining ability by using this LDS method. In order to make our data to qualify the minimum requirement of using the number count clustering analysis\footnote{To suppress the number count shot noise.}, we have to make sure that in each of the luminosity distance bins we have enough GW sources. Hence, we divide the whole range of luminosity distance from 0 to 30Gpc into three wide bins. Each of the bins contains comparable GW source numbers.  

For BBO, our results showed that the constraining ability is sensitive to the $D_L$ uncertainty. The Fisher forecasted cosmological parameter errors increases with increasing of the intrinsic GW luminosity measurement errors. The fractional errors on the cosmological parameters in the first luminosity distance bin, which is centered at 5Gpc, can achieve the level of a few $10^{-4}$ level under the circumstance of intrinsic $D_L$ measurement error of $\sigma^{\rm GW}_{\log D_L}=0.001$. When we degrade the intrinsic measurement error to the level of $\sigma^{\rm GW}_{\log D_L}=0.02$, the cosmological parameter errors also degrade about 1 order of magnitude. The cosmological parameter fractional errors in the second and third bins monotonically increase by a factor of a few compared with the previous adjacent bin. At the $D_L=25$Gpc distance, the fractional errors are about $10^{-1}$ level for all three parameters which are concerned. Once we include the velocity dispersion errors, the cosmological parameter fractional errors in the first bin are enlarged by roughly one order magnitude. However, in the second and third bins, the velocity dispersion has very limited impact on the final results. On the contrary, the lensing magnification errors modify the results mainly in the high redshift leaving the low redshift almost unchanged. For instance, even optimally assuming a $50\%$ delensing efficiency, the fractional errors on the cosmological parameters in the third luminosity distance bin are still degraded by one order magnitude compared with the one in the same bin but with velocity dispersion errors. Hence, we can not use the third bin to do any precision cosmology studies if we could not remove this lensing error significantly. The results for ET are similar as those from BBO. Due to the GW source number is less than the former, the corresponding results also get a bit worse. Moreover, because of the increment of the measurement shot noise, the cosmology results from ET are also less sensitive to the velocity dispersion and lensing errors. 

Ideally, the GW source number count can be used as a novel tracer of LSS in the luminosity distance space. In practice, once we include the peculiar velocity dispersion and lensing magnification uncertainties in $D_L$, we can only use the GW data as the ``photometric'' data instead of the originally proposed as the ``spectroscopic'' data. And due to the uncertainty along the line of sight, the robustness of this method is degraded significantly in the high redshift.

\section*{Acknowledgements}
QY is supported by the National Natural Science Foundation of China Grants No. 12005146. 
BH is supported by the National Natural Science Foundation of China Grants No. 11973016.


\bibliographystyle{mnras}
\bibliography{lds}



\bsp	
\label{lastpage}
\end{document}